\documentstyle[12pt]{article}
\begin{document}
\title{Supersymmetry of a spin $1\over 2$ particle on the real line}
  \author{V. M. Tkachuk, P. Roy$^{\dag}$\\ 
  {\small  Ivan Franko Lviv State University, 
         Chair of Theoretical Physics }\\ 
	{\small 12 Drahomanov Str., Lviv UA--290005, Ukraine}\\
	{\small E-mail: tkachuk@ktf.franko.lviv.ua}\\
 {\small $^{\dag}$ Permanent Address: Physics and Applied Mathematics 
 Unit,}\\ 
	{\small Indian Statistical Institute, Calcutta 700035, India}\\
	{\small E-mail: pinaki@isical.ac.in}}

\maketitle

\begin{abstract}
We study one dimensional supersymmetric (SUSY) quantum mechanics of a
spin $1\over 2$ particle moving in a rotating magnetic field 
and scalar potential. We also discuss
SUSY breaking and it is shown that SUSY breaking essentially
depends on the strength and period of the magnetic field.
For a purely rotating magnetic field the eigenvalue problem 
is solved exactly
and two band energy spectrum is found.


\end{abstract}

\section{Introduction}
Supersymmetric (SUSY) quantum mechanics was introduced by Witten
as a laboratory for investigating SUSY breaking which is one
of the fundamental issue in SUSY quantum field theory \cite{Wit}.
Prior to Witten's paper Nicolai had shown that SUSY could also be a useful
tool in nonrelativistic quantum mechanics \cite{Nic}.
Subsequently SUSY quantum mechanics has proved to be interesting 
on its own  
merit and has been studied from different points of view \cite{Coop,Jun}.

In the present paper we shall study a generalised one dimensional SUSY 
quantum mechanical problem
concerning the motion of a spin $1\over 2$ particle in the presence of 
a scalar potential
as well as a magnetic field. In this connection we would like to point out 
that supersymmetry based methods have previously 
been used to study various systems
involving coupled channel problems \cite{Ama},
matrix Hamiltonians \cite{Hau,And}
as well as models involving spin-orbit coupling \cite{Lev}. 
Quasi exactly solvable
matrix models have also been studied \cite{Zhd}.
In the present case we shall obtain
exact solutions of the eigenvalue problem when a spin $1\over2$
particle moves in the presence of a rotating magnetic field.
In particular it
will be shown that supersymmetry breaking depends nontrivially on the 
strength and period
of the magnetic field. Finally we shall also indicate briefly 
how supersymmetry
is affected when apart from the rotative 
magnetic field a scalar potential is also present.

\section{SUSY of a spin $1\over 2$ particle on the real line}

In Witten's model of SUSY quantum mechanics the Hamiltonian
consists of two factorized Schr\"{o}dinger operators

\begin{equation}
H_-=A^+A^-,~~~~~ H_+=A^-A^+,
\end{equation}
where the operators $A^+$ and $A^-$ are given by

\begin{equation}
A^{\pm} = \mp\frac{d}{dz} + W(z),
\end{equation}
and $W(z)$ is the superpotential.

The pair of Hamiltonians in (1) are called SUSY partner Hamiltonians and 
each of these Hamiltonians describe motion of a spinless particle in 
one dimension. 
We shall now generalize Witten's model of SUSY quantum mechanics in such way
that each of the Hamiltonians $H_-$, $H_+$ will describe the motion 
of a spin 
$1\over 2$
particle in a magnetic field and scalar potential. 
In order to do this we generalise the operators $A^{\pm}$ in the following 
way: 
\begin{equation}
A^{\pm}=\mp{d\over dz}+W(z)+{\bf V}(z){\bf S},
\end{equation}
It may be noted that here we consider motion of the particle along 
$z$-axis and
components of the spin operator ${\bf S}$ are 
$S_{\alpha}=\sigma_{\alpha}/2$ ($\alpha=x, y, z$),
$\sigma_{\alpha}$ being the Pauli matrices. Then SUSY partner 
Hamiltonians can be
obtained as in (1) and are given by
\begin{equation}\label{Hpm}
H_{\pm}=-{d^2\over{dz^2}}+V_{\pm}(z)+{\bf B_{\pm}}(z){\bf S},
\end{equation}
where
\begin{eqnarray}
V_\pm (z)=W^2 \pm W'+V^2/4,\\
{\bf B}_\pm (z)=2W{\bf V}\pm{\bf V'}.
\end{eqnarray}
The Hamiltonians $H_\pm$ in (\ref{Hpm}) describe a spin $1\over 2$ 
particle moving along the
$z$-axis in a scalar potential $V_\pm$ and a magnetic field 
${\bf B}_\pm (z)$.
In the case $ {\bf V}=0$ we obtain standard Witten model of SUSY quantum 
mechanics.

\section{Spin $1\over 2$ particle in a rotating magnetic field 
and constant scalar potential}
Let us now consider the motion of spin $1\over 2$ particle in a constant 
scalar potential and 
rotating magnetic field in the $x-y$-plane:
\begin{equation}
(B_\pm)_x={\mp}B_0 \cos kz, \ \
(B_\pm)_y={\mp}B_0 \sin kz, \ \
(B_\pm)_z=0.
\end{equation}
In this case without any loss of generality we can choose $W=0$ and thus
\begin{equation}
V_x=-{B_0\over k} \sin kz, \ \
V_y={B_0\over k} \cos kz, \ \
V_z=0.
\end{equation}
Thus in this case the operators $A^{\pm}$ are given by
\begin{equation}
A^{\pm}=\mp{d\over dz}+{B_0\over k}(-\sin kz S_x +\cos kz S_y).
\end{equation}
Then from (4) we can obtain the explicit form of the  
SUSY partner Hamiltonians: 
\begin{equation} \label{H0}
H_\pm=-{d^2\over dz^2} \mp B_0(\cos kz S_x +\sin kz S_y)
+{B_0^2\over 4k^2}.
\end{equation}

From the form of the Hamiltonians in (10) it is seen that the term 
coupling spin and magnetic field is 
dependent on z. In order to remove this dependence we now perform the 
following unitary transformation:
\begin{equation} \label{Utr}
{\tilde \psi}=e^{ikzS_z}\psi.
\end{equation}

As a result of this transformation we obtain the following 
set of Hamiltonians 
\begin{eqnarray} \label{H0t}
{\tilde H_\pm}=e^{ikzS_z}He^{-ikzS_z}={\tilde A^{\mp}}{\tilde A^{\pm}} 
\nonumber \\
=\left(-i{d\over dz} - kS_z \right)^2 \mp B_0 S_x +{B_0^2\over 4k^2},
\end{eqnarray}
where
\begin{equation}
{\tilde A^\pm}=e^{ikzS_z}A^\pm e^{-ikzS_z}=
\mp{d\over dz}\pm ikS_z+{B_0\over k}S_y.
\end{equation}

In order to determine whether or not supersymmetry is broken it is
necessary to investigate if there are normalisable zero energy ground
state wave functions (it may be recalled that for SUSY to be unbroken 
the ground state energy must be zero while if
SUSY is broken the ground state energy is positive). So we seek solutions
of the equations
\begin{equation}
{\tilde A^{\pm}}{\tilde \psi}_0^{\pm} = 
\left(\mp{d\over dz}\pm ikS_z+{B_0\over k}S_y\right){\tilde \psi}^\pm_0=0.
\end{equation}
Thus SUSY is unbroken if at least one of wave functions $\tilde \psi_0^\pm$
is a true zero mode. We now seek solutions of the above equations 
in the form
\begin{equation}
{\tilde \psi_0^\pm}={\tilde \chi}_0^\pm e^{iqz},
\end{equation}
where $q$ is wave vector of the particle, ${\tilde \chi}_0^\pm$ is spin
part of the wave function and it satisfies the following equation
\begin{equation}
\left(\mp iq \pm ikS_z+{B_0\over k}S_y\right){\tilde \chi}_0^\pm=0.
\end{equation}
It can be shown that non zero solutions of equation (16) i.e., 
${\tilde \chi}_0^+$ or
${\tilde \chi}_0^-$ exists for the same value of the wave vector $q$
\begin{equation} \label{q0}
q=\pm {k\over 2}\sqrt{1-{B_0^2\over k^4}}.
\end{equation}
This implies that true zero modes exist both for $H_+$ and $H_-$ only when
$q$ is real and from (17) it follows that $q$ is real if
\begin{equation} \label{ExSUSY}
{B_0^2\over k^4}< 1.
\end{equation}
Thus when $B_0^2 < k^4$ zero modes exist both for $H_{\pm}$ and SUSY 
is unbroken.
In the other case when q is complex we do not have normalisable zero energy 
solutions and so
SUSY is broken.
It is interesting to note that when q is real zero energy states
exist in both the sectors $H_+$ and $H_-$ and thus are strictly isospectral.
It may be noted that a similar situation arises when spinless particles
move in periodic potentials \cite{DunF,DunM}.

Finally we note that eigenvalue problem for the Hamiltonians in 
(\ref{H0t}) and
thus for those in (\ref{H0}) can be solved exactly for the entire 
energy energy spectrum.
After performing the unitary transformation 
the Hamiltonian (\ref{H0}) is transformed to (\ref{H0t})
and the corresponding eigenfunctions can be written in the form
\begin{equation}
{\tilde \psi^\pm}={\tilde \chi}^\pm e^{iqz}.
\end{equation}
Then the eigenvalue problem becomes
\begin{equation}
[(q-kS_z)^2\mp B_0 S_x]{\tilde \chi}^\pm =E{\tilde \chi}^\pm,
\end{equation}
from which we obtain a two band energy spectrum
\begin{equation} \label{Eq}
E_{1,2}(q)=q^2+(k/2)^2\pm\sqrt{q^2k^2+(B_0/2)^2}+{B_0^2\over 4k^2}.
\end{equation}
We note that the energy spectrum in (\ref{Eq}) is the same for both $H_\pm$.
The lowest energy $E=0$ for the first band $E_1(q)$ 
(with "-" in (\ref{Eq})) is at wave vector given by (\ref{q0})
where $B_0^2/k^4<1$.
If however $B_0^2/k^4>1$ the lowest energy is at $q=0$ and
$E_1(q = 0) = (k/2-|B_0|/2k)^2$ which is greater than zero. 
Thus in this case the SUSY is broken. 

\section{Ground state in the case of a rotating magnetic field and
non constant scalar potential}
Unlike in the last section here we consider the motion of the particle in
a rotating magnetic field and a non constant superpotential $W(z)$. In this
case the equation for the ground state of $H_-$, after 
the unitary transformation (\ref{Utr}) is given by
\begin{equation} \label{EpsiW}
{\tilde A}^-{\tilde \psi}_0^- =
\left({d\over dz}- ikS_z+{B_0\over k}S_y+W(z)
\right){\tilde \psi}_0^-=0.
\end{equation}
As before spin and coordinate parts of the wave function can be 
separated and 
the solution can be written in the form 
\begin{equation} \label{psiW}
{\tilde \psi}^-_0={\tilde\chi}^-_0 \exp(-\int^z W(z)dz-\lambda z)
\end{equation}
where ${\tilde\chi}^-_0$ satisfies the equation
\begin{equation} \label{chi}
\left(- ikS_z+{B_0\over k}S_y
\right){\tilde \chi}^-_0=\lambda {\tilde\chi}^-_0.
\end{equation}
Eigenvalues of this equation are easily obtained and are given by
\begin{equation} \label{lamb}
\lambda =\pm{k\over2}\sqrt{{B_0^2\over k^4}-1}.
\end{equation}
Thus equation (\ref{EpsiW}) has two solutions (\ref{psiW})
which correspond to the two eigenvalues (\ref{lamb}).

An interesting feature which emerges from this scenario is that 
even in the presence of a non constant scalar potential
the rotating magnetic field can lead to SUSY breaking
if it is sufficiently strong.
To see this let us choose a 
superpotential $W(z)$ such that $W(z)\rightarrow \pm W_0$ when 
$z\rightarrow \pm \infty$.
Then it follows from
(\ref{psiW}) that in the case when
\begin{equation} \label{BrSUSY}
{k\over2}\sqrt{{B_0^2\over k^4}-1}>W_0
\end{equation}
the wave function becomes non square integrable. Thus sufficiently strong 
magnetic field destroys zero energy ground state
and leads to SUSY breaking.

For the purpose of illustration let us consider an explicit example.
We choose 
$W(z)=\alpha \tanh z$, $\alpha >0$ so that $W(z) 
\rightarrow \pm \alpha$ as $z \rightarrow \pm \infty$.
Then the ground state wave function corresponding to  $H_-$ is given by
\begin{equation}
{\tilde\psi}^-_0={\tilde\chi}^-_0(\cosh z)^{-\alpha}e^{-\lambda z},
\end{equation}
where $\lambda$ is given by (\ref{lamb}). This wave function
is square integrable, when $\alpha >\lambda$.
Thus in this case SUSY is unbroken. In the other case when
$\alpha <\lambda$ the ground state wave function is nonnormalisable 
so that the
magnetic field leads to the breaking of SUSY.
We would like to point out that if $\lambda$ as given by (\ref{lamb})
is imaginary (in other words if the magnetic field is small enough)
then SUSY is always unbroken irrespective of the value of $\alpha$. 

Finally let us point out about the zero energy solution corresponding to 
$H_+$. We note that in this case 
\begin{equation}
{\tilde\psi}^+_0={\tilde\chi}^+_0(\cosh z)^{\alpha}e^{\lambda z},
\end{equation}
so that it is non square integrable. Thus $H_+$ posses no zero
energy solution. This is in contrast to the case considered in the
previous section where both $H_\pm$ had zero energy states.

\section{Conclusions}
In the present paper we have studied motion of a spin $1\over 2$ particle in
rotating magnetic field and a scalar potential within the framework 
of SUSY quantum mechanics.
The eigenvalue problem in the case of a purely magnetic field 
(scalar potential is constant) is solved exactly and two band energy 
spectrum is obtained.
An interesting feature of the free motion of spin $1\over 2$ particle in a
rotating magnetic 
field is that both Hamiltonians $H_+$ and $H_-$ can have zero energy states
simultaneously. It may be noted that existence of zero modes and thus 
exact SUSY depends on
parameters of magnetic field and are given by (\ref{ExSUSY}).
For sufficiently strong magnetic field we have broken SUSY.

We have also studied SUSY breaking when the particle is moving 
in a rotating
magnetic field and non constant superpotential $W(z)$. 
In contrast to the free motion ($W(z)=0$) for non constant 
$W(z)$ only one of the Hamiltonians
$H_-$ or $H_+$ has a zero energy ground state. 
In this case if the magnetic field
is sufficiently strong then SUSY can be broken.
The condition for this is given by (\ref{BrSUSY}).

It may be noted that 
in addition to the magnetic field inclusion of a 
non constant superpotential $W(z)$ leads to the appearance of discrete 
energy levels.
Investigation of complete discrete energy spectrum in the presence of a
magnetic field and different scalar potentials will 
be the subject of a future publication.


\end{document}